\documentstyle[12pt,epsf]{article}
\begin{document}

\begin{center}
{\Large\bf  SOLUTION OF THE ODDERON PROBLEM  
    FOR ARBITRARY CONFORMAL WEIGHTS                }\\
\vspace*{2cm}
{\large\bf J. Wosiek and R. A. Janik}  \\
\mbox{ }\\
Institute of Physics, Jagellonian University \\
Reymonta 4. 30-059 Cracow, Poland \\
\end{center}
\vspace*{2cm}

\begin{abstract}
 A new method is applied to solve the Baxter equation for 
three coupled, noncompact spins. 
 Due to the equivalence  with the system 
of three reggeized gluons,  
the intercept of the odderon trajectory is
predicted for the first time, as the analytic function of the two relevant
parameters.

\end{abstract}
 
\vspace*{2cm}
\noindent TPJU-21/96 \newline
October 1996 \newline
hep-th/9610208
\newpage

Calculation of the QCD prediction for the intercept of the odderon
trajectory still remains a challenge for the Leading Logarithmic
 scheme of the reggeization
of QCD \cite{firsta,firstb}. 
In the first approximation the problem naturally separates into sectors with 
fixed number $n$ of the reggeized gluons propagating in the $t$ channel.
The lowest nontrivial case, $n=2$, was solved in the classical papers 
by Balitskii,
Kuraev, Fadin and Lipatov \cite{BFKL}
resulting in the simple expression for the intercept of the hard pomeron.
The notable progress for arbitrary $n$ was achieved by Lipatov and 
Faddeev and Korchemsky  
\cite{LIP0,FK} who have established
exact equivalence with the one dimensional chain of $n$ noncompact spins.
Leading high energy behaviour of QCD amplitudes is given by the highest 
eigenvalue of the
corresponding Heisenberg hamiltonian of $n$ spins with nearest
neighbour interaction. Moreover, by identifying enough constants of motion
they were able to prove that this system is soluble for arbitrary $n$. 
The success of this, rather mathematical, approach was confirmed by 
rederiving
the Lipatov et al. result in the $n=2$ case \cite{FK,kor1}.
 However, the adopted procedure 
requires an analytic continuation from the integer values
of the relevant conformal weight $h$ (see later) because only for integer $h$
they were  able to diagonalize the two spin hamiltonian.  The $n=3$ case,
 which gives the lowest contribution to the 
odderon exchange, was also studied by  Faddeev and Korchemsky, 
and Korchemsky \cite{FK,kor1}.
Again, the spectrum of the system for integer $h$ can be found for any finite 
$h=m$. However, the general expression for arbitrary $m$ 
is not known, and
consequently the analytical continuation to $h=1/2$ is not available
\footnote{The lowest state of the $n=3$ hamiltonian is believed to occur
 at $h=1/2$.}.

 We have developed a new approach which a) works for arbitrary values
of the conformal weight $h$, 
providing explicitly above continuation,
and b) gives  the analytic solution of the $n=3$ case for arbitrary
$h$ and $q_3$. For $n=2$ our method reproduces again the $BFKL$ result
clarifying the problem of boundary conditions for arbitrary $h$. 
This and the details of the
$n=3$ calculation will be presented elsewhere \cite{webig}. 
In this letter we address
directly the odderon case and discuss some phenomenological consequences.  
 
The intercept of the odderon trajectory is given by 
\begin{equation}
\alpha_O(0)=1+{\alpha_s N_c \over 4\pi}\left(\epsilon_3(h,q_3)+
 \overline{\epsilon}_3(\overline{h},\overline{q}_3)\right), \label{inter}
\end{equation}
where $\epsilon_3$ and $\overline{\epsilon}_3 $ are respectively the largest 
eigenvalues
of the $n=3$ reggeon hamiltonian and its antiholomorphic counterpart
 \cite{FK,kor1}. This system is equivalent to the 
one dimensional chain
of three noncompact spins with nearest-neighbour interactions. Applying
 Bethe ansatz to the latter one obtains
\begin{equation}
\epsilon_3=i \left({\dot{Q}_3(-i)\over Q_3(-i)}-{\dot{Q}_3(i)\over Q_3(i)} 
                            \right)-6,
\end{equation}
where $Q_3(\lambda)$ satisfies the following Baxter equation
\begin{equation}
(\lambda+i)^3 Q_3(\lambda+i)+(\lambda-i)^3 Q_3(\lambda-i)=
(2\lambda^3+q_2\lambda+q_3) Q_3(\lambda).  \label{bax}
\end{equation}
$q_2$ and $q_3$ are the eigenvalues of the two, commuting with the
hamiltonian, operators which play important role 
in the proof of the solubility of the above system
\cite{LIP0,FK}. 
The spectrum of $\hat{q}_2$ is known from the symmetry considerations
\begin{equation}
q_2=h(1-h),\;\;\;\;h={1\over 2}(1+m) -i\nu,\;\;  m\in Z, \nu\in R.
\end{equation}
The eigenvalues of $\hat{q}_3$ are known only for integer conformal
weights $h$, whereas the value of $q_3$ for the ground state
of the three reggeized gluons $(h=1/2)$ is not available.
Analogous expressions hold for the antiholomorphic sector
with $\overline{h}=(1-m)/2-i\nu$ \cite{kor1}. 

 Our goal is to determine $\epsilon_3(h,q_3)$ for arbitrary $h$ and $q_3$.
  To this end we begin with the trick of Ref.\cite{jan1} and seek
  the solution of the Baxter equation (\ref{bax}) in the form of the
  {\em double} contour representation 
\begin{equation}
Q_3(\lambda)=\int_{C_I} Q_I(z)  z^{-i\lambda-1} (1-z)^{i\lambda+1} dz
            + \int_{C_{II}} Q_{II}(z)  z^{-i\lambda-1} (1-z)^{i\lambda+1} dz.
            \label{con}
\end{equation} 
Provided the boundary contributions cancel,
 Eq.(\ref{bax}) is equivalent to the following ordinary differential
 equation for $Q_I(z)$ and $Q_{II}(z)$  $(\equiv Q(z))$
 \begin{equation}
 \left[ \left(z(1-z){d\over dz}\right)^3 -q_2z^2(1-z)^2 {d\over dz} 
 +i q_3 z(1-z)
 \right] Q(z)=0. \label{diff}
\end{equation}
This is a third order linear equation of the Fuchs class
with the three regular singular points at $z=0,1$ and $\infty$, considered 
earlier in Refs. \cite{kor1},\cite{Wallon},\cite{Melb}
We will prove that the complete boundary conditions on $Q_I(z)$ and $Q_{II}(z)$
 are {\em uniquely }
determined by the requirement of the cancellation of the boundary
terms among the integrals (\ref{con}). This is the distinctive feature
of the $n=3$ case which allows for the successful application of our strategy. 

We begin with the construction of the two fundamental sets of three,
linearly independent  solutions
\begin{eqnarray}
\vec{u}(z) = (u_1(z),u_2(z),u_3(z)), \\ \nonumber 
\vec{v}(z) = (v_1(z),v_2(z),v_3(z)),\\ \nonumber
\end{eqnarray}
around $z=0$ and $z=1$ respectively.   
\begin{eqnarray}
u_1(z)&=& \sum_{n=0}^{\infty} f_n z^n, \nonumber \\
u_2(z)&=& {1\over \pi i}\log{z}\; u_1(z) + 
{1\over \pi i}\sum_{n=0}^{\infty} r^{(1)}_n z^n, \label{uba} \\ 
u_3(z)&=& {1\over \pi^2}\log^2{z}\; u_1(z)+ 
          {2\over\pi^2}\log{z}\; \sum_{n=0}^{\infty} r^{(1)}_n z^n
          + {1\over \pi^2}\sum_{n=0}^{\infty} r^{(2)}_n z^n, \nonumber   
\end{eqnarray} 
where the coefficients of the expansions are determined by the recursion
relations easily obtained from Eq.(\ref{diff}). 
\begin{eqnarray}
f_{n+1}&=&(b_n f_n-c_{n-1}f_{n-1})/a_{n+1},\;\; f_0=1,\;f_{-1}=0,\nonumber \\
a_n&=&n^3,\nonumber \\ 
b_n&=&iq_3+n(q_2+(2n+1)(n+1)), \label{rec}\\
c_n&=&n(q_2+(n+1)(n+2)), \nonumber
\end{eqnarray}
and for the logarithmic solutions
\begin{eqnarray}  
r^{(1)}_{n+1}&=&(-p^{(1)}_n+b_n r^{(1)}_n-c_{n-1} r^{(1)}_{n-1})/a_{n+1},
\;\;r^{(1)}_0=1,\;r^{(2)}_{-1}=0,\nonumber \\
p^{(1)}_n&=&3(n+1)^2f_{n+1}-(1+q_2+6n(n+1))f_n\\
         &+&(-1+q_2+3n^2)f_{n-1},\nonumber \\
r^{(2)}_{n+1}&=&(-p^{(2)}_n+b_n r^{(2)}_n-c_{n-1} r^{(2)}_{n-1})/a_{n+1},
\;\;r^{(2)}_0=1,\;r^{(2)}_{-1}=0,\nonumber \\
p^{(2)}_n&=&6(n+1)f_{n+1}-6(2n+1)f_n+6nf_{n-1}\\
    &+&6(n+1)^2r^{(1)}_{n+1}-2(1+q_2+6n(n+1))r^{(1)}_n\\
    &+&2(2+q_2+3(n^2-1))r_{n-1} .
  \label{recl}
\end{eqnarray}
The series in Eq.(\ref{uba}) are convergent in the unit circle $K_0$ around
$z=0$. Similarly one can construct the $\vec{v}(z)$ solutions in the unit
circle $K_1$ around $z=1$. In fact, because of the 
symmetry of the Eq.(\ref{diff})
 we take
\begin{equation}
\vec{v}(z;q_2,q_3)=\vec{u}(1-z;q_2,-q_3), \label{vba}
\end{equation}
Since any solution is the
linear combination of the fundamental solutions, we have
\begin{eqnarray}
Q_I(z)=a u_1(z)+b u_2(z) + c u_3(z) \equiv
      A\cdot\vec{u}(z)=A\cdot \Omega \vec{v}(z),\nonumber \\
Q_{II}(z)=d u_1(z)+e u_2(z) + f u_3(z) \equiv
      B\cdot\vec{u}(z)=B\cdot \Omega \vec{v}(z), 
      \label{abf}
\end{eqnarray}
with an obvious vector notation.   
The transition matrix $\Omega$ is defined by
\begin{equation}
\vec{u}(z)=\Omega \vec{v}(z), \label{trans}
\end{equation}
and  plays an important role in the following.
It provides the analytic continuation of our solutions $Q(z)$ between
$K_0$ and $K_1$. Transition matrix 
 $\Omega$ can be easily determined from Eq.(\ref{trans}) once 
 both bases, Eqs.(\ref{uba},\ref{vba}), are known. For example,
 the i-th row, $\vec{\omega}_i^T$, of $\Omega$ can be obtained as
\begin{equation}
\vec{\omega}_i=(\Sigma^T)^{-1}\vec{w}_i, \;\;\; \Sigma_{kr}=v_k(z_r),\;\;
(\vec{w_i})_r=u_i(z_r),\;\;\; i,k,r=1,2,3.
\end{equation}
Where $z_1,z_2$ and $z_3$ are arbitrary three points inside the intersection
of $K_0$ and $K_1$. Next we introduce the monodromy matrix $M$
which describes the behaviour of the basis $\vec{u}$ in the vicinity of
 the branch point $z=0$ (see Fig.1)
\begin{equation}
\vec{u}(z_{end})=M \vec{u} (z_{start}), \;\;
M=\left( \begin{array}{ccc}
                 1&0&0 \\
                 2&1&0\\
                -4&-4&1 
         \end{array} \right) . \label{mon} 
\end{equation}                                 
 
We are now ready to write the condition for the cancellation of the
boundary contributions in Eq.(\ref{con}). 
 Let us choose the contours $C_I$
and $C_{II}$ as shown in Fig.2. Then, the boundary contributions cancel if
\begin{equation}
  A^T M_I = P^T , \;\;\;  B^T M_{II}= -P^T,\;\;\;  P^T=(\alpha, \beta, \gamma),
  \label{canc}
\end{equation}
where the combined monodromy matrices for the corresponding contours
read
\begin{equation}   
M_I=\Omega M \Omega^{-1} - M^{-1},  \;\;
M_{II}=\Omega M^{-1} \Omega^{-1} - M. \label{com}
\end{equation}
Hence the original freedom of six coefficients in Eqs.(\ref{abf}) was reduced
to the three free parameters which we conveniently
choose as $\alpha,\beta$ and $\gamma$. This was expected. However,
in the n=3 case additional simplification occurs which, remarkably,
 allows to fix completely the remaining freedom. 
 
 The key point is the observation that the monodromy matrices (\ref{com})
 are singular, i.e. $det(M_I)=det(M_{II})=0$. To see this it is enough
 to inspect the Riemann P symbol corresponding to Eq.(\ref{diff}).
 \begin{equation}
 P\left\{    \begin{array}{ccc}
              0 & 1 & \infty  \\
              0 & 0 &  0      \\
              0 & 0 & 1+h     \\
              0 & 0 & 2-h     \\
              \end{array}; z \right\}
 \end{equation}
 It is readily seen that, contrary to the n=2 case, there exists 
 a solution which is {\em regular} at $z=\infty$. Therefore
 the reduced monodromy matrices $M_I$ and $M_{II}$ must have the zero eigenvalue
 corresponding to this solution. As a consequence, Eqs.(\ref{canc})
 do not have the unique solution for (a,...,f). Different choices
 of coefficients differ by the zero mode. This difference
 is inessential because the integrals of the solution, regular at infinity,
  vanish. We therefore proceed
 to isolate the zero mode explicitly, and then impose the condition
  of cancellation of boundary contributions. To this end we introduce 
  the new basis $\vec{t}(z)$ such that $t_3(z)$ is the solution regular at
  $z=\infty$. Transformation matrix ${\cal T} $, $\vec{u}(z)={\cal T} \vec{t}(z)$,
  to the $\vec{t}(z)$ basis can be readily
  obtained by diagonalizing the commuting matrices
  $M^A=M M_I$ and $M^B=M_{II} M^{-1}$. In the new basis (marked by the 
subscript $t$) 
the cancellation
  condition reads
  \begin{equation}
  A_t'^T M_t^A = P_t^T,\;\;\; B_t^T M_t^B=-P_t^T M_t^{-1}, \label{cann}
  \end{equation}
  where $M_t^A$ and $M_t^B$ are diagonal with one (third, say) zero eigenvalue,
  $A_t'^T\equiv (a_t',b_t',c_t')=A_t^T M^{-1}_t,$ and 
$M_t={\cal T}^{-1} M {\cal T}$ 
  is the monodromy
   matrix in the new basis. Our final conclusions follow now trivially
   from Eq.(\ref{cann}). Coefficients $c_t'$ and $f_t$ are arbitrary
 (and irrelevant), $\gamma_t=0$ and $a_t',b_t',d_t,e_t$ are determined uniquely
 by $\alpha_t$ and $\beta_t$ provided the {\em additional consistency condition}
 \begin{equation}
 \alpha_t m_{13}+\beta_t m_{23} =0,  \label{newc}
 \end{equation}
 is satisfied, $m_{ik}$ being the matrix elements of $M_t^{-1}$.
 The last condition fixes completely the remaining freedom. 
 Now the integral transforms $Q_I(z)$ and $Q_{II}(z)$ 
 (hence also the solution of the Baxter equation $Q(\lambda) )$ are determined
 uniquely up to an irrelevant normalization. This ends our proof. 
 
 Now the contour integrals and  derivatives over $\lambda$
  can be done analytically since
 $C_I$ and $C_{II}$ lay within the corresponding domains of convergence
 of all involved series \footnote{We use the $u (v)$ basis on $C_I (C_{II})$.}.
 Integrating resulting expressions term by term \footnote{Consistent choice
of the appropriate branches of the kernel and of the multivalued
functions $Q_{I,II}(z)$ must be made \cite{webig}.}
we have obtained the final
 formula for $\epsilon_3(h,q_3)$ in the form of
 absolutely convergent series for arbitrary values of the conformal
 weight $h$ and $q_3$. Resulting expression is rather lengthy and will not
 be quoted here, however it provides, for the first time, the energy of the
  three reggeon hamiltonian as the analytic function of the relevant parameters. 
  
  We will now discuss some special cases and phenomenological consequences
  of our result. First,  for integer $h=m$ there exists a discrete set
  of quantized values of $q_3=q_3^k(m)$ for which the polynomial solution
  $u_1(z)$ exists. This quantization of $q_3$ is known and the eigenenergies
  $\epsilon_3 (\overline{\epsilon_3}) $ at these points can be
   calculated \cite{kor1}. We quote the first few levels in Table 1.

  \begin{table}    
  \begin{center}
   \begin{tabular}{|c|cc|cc|} \hline\hline
   {\em h} & \multicolumn{2}{c|}{ $q_3$ } 
                                     & \multicolumn{2}{c|}{ $\epsilon_3$} \\
   \hline
  4 & \multicolumn{2}{c|}{$\pm 2\sqrt{3}$}
                                     & \multicolumn{2}{c|}{$-7{1\over 2}$} \\
  5 & \multicolumn{2}{c|}{$\pm 6\sqrt{3}$} 
                                     & \multicolumn{2}{c|}{$-8{5\over 6}$} \\
  6 & $\pm 4\sqrt{3}$ & $\pm 4\sqrt{30}$ & $-9{1\over 4}$  & $ -10 $  \\ 
   \hline\hline  
   \end{tabular}
  \end{center}
\caption{Quantization of $q_3$ and first few levels of the holomorphic
hamiltonian in the polynomial case.}
   \end{table}

   Our formula reproduces these results exactly. The whole procedure
   can be followed in the polynomial case and it can be shown that the known
   spectrum emerges. 
 Two particular cases are shown in Figs. 3a and 3b. 
   In Fig.3a we plot $\epsilon_3(h, 2\sqrt{3})$ around $h=4$, and 
   in Fig.3b $\epsilon_3(6,q_3)$ is shown in the region of $q_3$ which contains
   both positive eigenvalues $q^{(1,2)}_3(6)$ listed in Table I. All levels
   of $\epsilon_3$ (c.f. Table I) are reproduced. 
   Second, our expression agrees with the asymptotic formula derived in 
   Ref.\cite{kor2}. In the limit $h, q_3\rightarrow\infty,\;\; q_3/h^3=const.$,
   Korchemsky has derived a simple  expression
   \begin{equation}
   \epsilon_3(h,q_3)=-2\log{2}-\sum_{k=0}^{3} \left[ \psi(1+i h x_k )
   +\psi(1-i h x_k) -2\psi(1) \right],  
      \label{asym}
   \end{equation}
   where $x_k,\; k=1,2,3$ are three roots of the polynomial
   $ 2 h^3 x^3+h^2(1-h) x+ q_3 $. 
   Fig. 3c shows comparison of this asymptotic form with our exact formula
  for $q_3/h^3=1$. Agreement is very striking indeed and persists to $h$ as low
   as $h\sim 0.4$ .  Interestingly, it turns out that the expression 
  (\ref{asym}) contains many terms which are nonleading in the above limit.
  Retaining consistently the terms up to a given order in $1/h$,
  as was also done in Ref.\cite{kor2}, gives yet simpler result which however
  does not work so well. Also the analytic structure of $\epsilon_3(h,q_3)$
  is reasonably well reproduced by Eq.(\ref{asym}) while the rigorous
  expansion in $1/h$ fails here (see later).
        
        Finally we discuss phenomenological consequences and relation  
with other works.
 Since the lowest state of the three reggeized gluons is expected to occur
 at $h=1/2$,
 we have mapped numerically the analytic structure of
 $\epsilon_3(1/2,q_3)$ in the complex $q_3$ plane. Results are sketched 
in Fig.4.  The holomorphic energy
  $\epsilon_3(1/2,q_3)$ has a series of simple poles on the imaginary axis,
  and behaves regularly in the remaining part of the $q_3$ plane. In fact
  Re $\epsilon_3(1/2,q_3)$ is negative almost in the whole $q_3$ plane
  except of the small regions in the vicinity of above poles.
  Interestingly, the approximate solution, Eq.(\ref{asym}), has the same
  singularity structure with similar location of poles. 
  This suggests that there may exist a better, than $1/h$ expansion,
   approximation scheme in which  Eq.(\ref{asym}) is the lowest order.
  
  Because of Eq.(\ref{inter}) and the symmetry 
  $(\overline{\epsilon}_n=\epsilon^{\star}_n)$, the intercept of the
  odderon trajectory is smaller than $1$ for most values of $q_3$.
  In particular, $\alpha_O(0) < 1$ for all real $q_3$. At the origin 
  \begin{equation}
  \epsilon_3(1/2,0)=-.738\dots ,  \label{zero}
  \end{equation}
  which indicates that the arguments used in  Ref.\cite{brau} 
  for the odderon intercept being exactly one, are incomplete.
  This discrepancy is caused by the degeneracy of the Baxter equations for 
  two and three reggeized gluons when $q_3=0$.  Continuity
 argument indicates that
  (\ref{zero}) is the true eigenvalue in the three reggeon sector.
 
  A variational estimate of the lower bound for the
  odderon intercept $\alpha_O(0) > 1+0.28g_s^2/\pi^2$ was derived in 
  Ref.\cite{lip3}.
  Together with present result it limits rather severely the allowed region of
  $q_3$ for the ground state of the system.
  
  Due to the singularity structure seen above,
  the final prediction for $\alpha_O(0)$ requires however more detailed 
  knowledge of the spectrum of $\hat{q}_3$. 
  Some progress in this area
  has been reported in Refs.\cite{kornew}, \cite{jan2}.    
  
  Our approach may be  generalized to higher $n$ \cite{webig}. Such a
  programme would provide the leading intercept of the $n$ reggeized gluons
  as the analytic function of the $n-1$ parameters $q_2,\dots ,q_n$. 
  
\vspace*{1cm}
  We thank L. Lipatov and G. Korchemsky for interesting discussions.
  This work is  supported by the Polish Committee for Scientific Research 
  under grants no. PB 2P03B19609 and PB 2P03B08308.

\section*{Figures}
\begin{figure}[htb]
\vspace{9pt}
\framebox[100mm]{
\epsfxsize=6cm \epsfbox{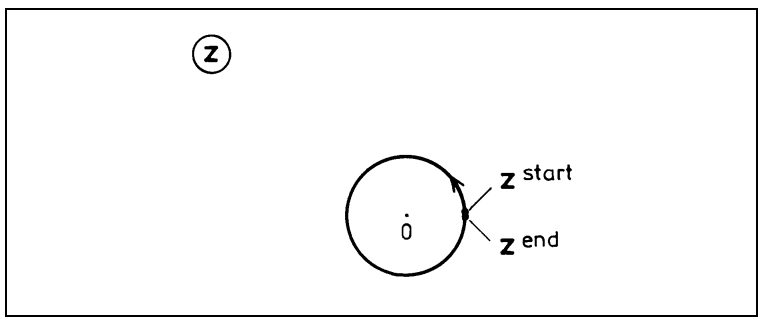}}
\caption{ Closed contour used to define the monodromy matrix, 
Eq.(\protect\ref{mon}). $z^{start}
=z^{end}$, however they belong to the different sheets of the Riemann surface.
}
\label{fig:f1}
\end{figure}

\begin{figure}[htb]
\vspace{9pt}
\framebox[100mm]{
\epsfxsize=9cm \epsfbox{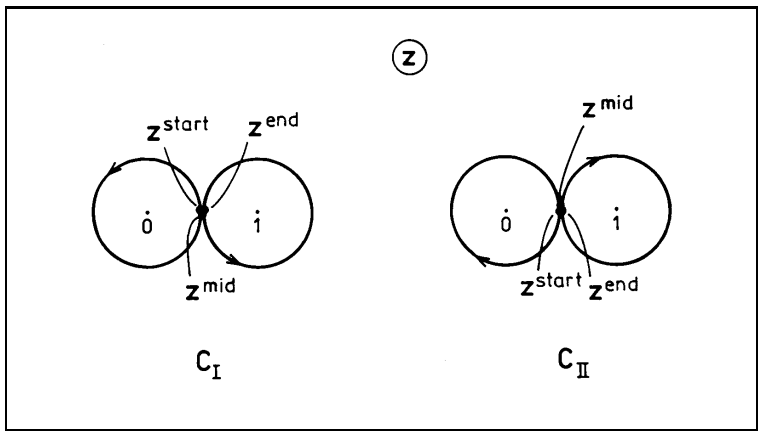}  }
\caption{ Integration contours used in Eq.(\protect\ref{con}). Start $z^{start}$,
middle $z^{mid}$, and end $z^{end}$ points coincide but they lay on the
different sheets of the Riemann surface of the integrands.
}
\label{fig:f2}
\end{figure}

\begin{figure}[htb]
\vspace{9pt}
\epsfxsize=7cm \epsfbox{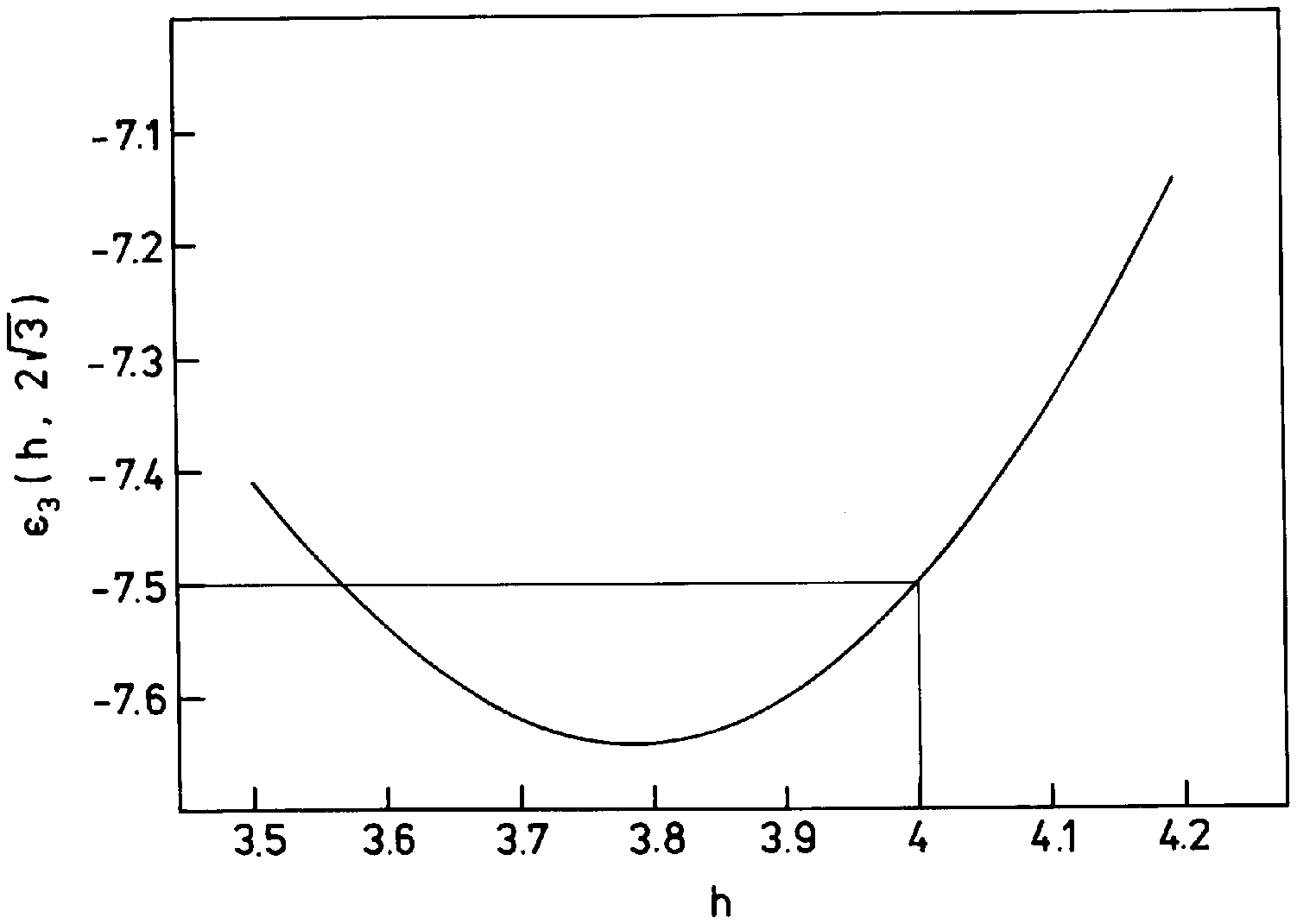}
\epsfxsize=7cm \epsfbox{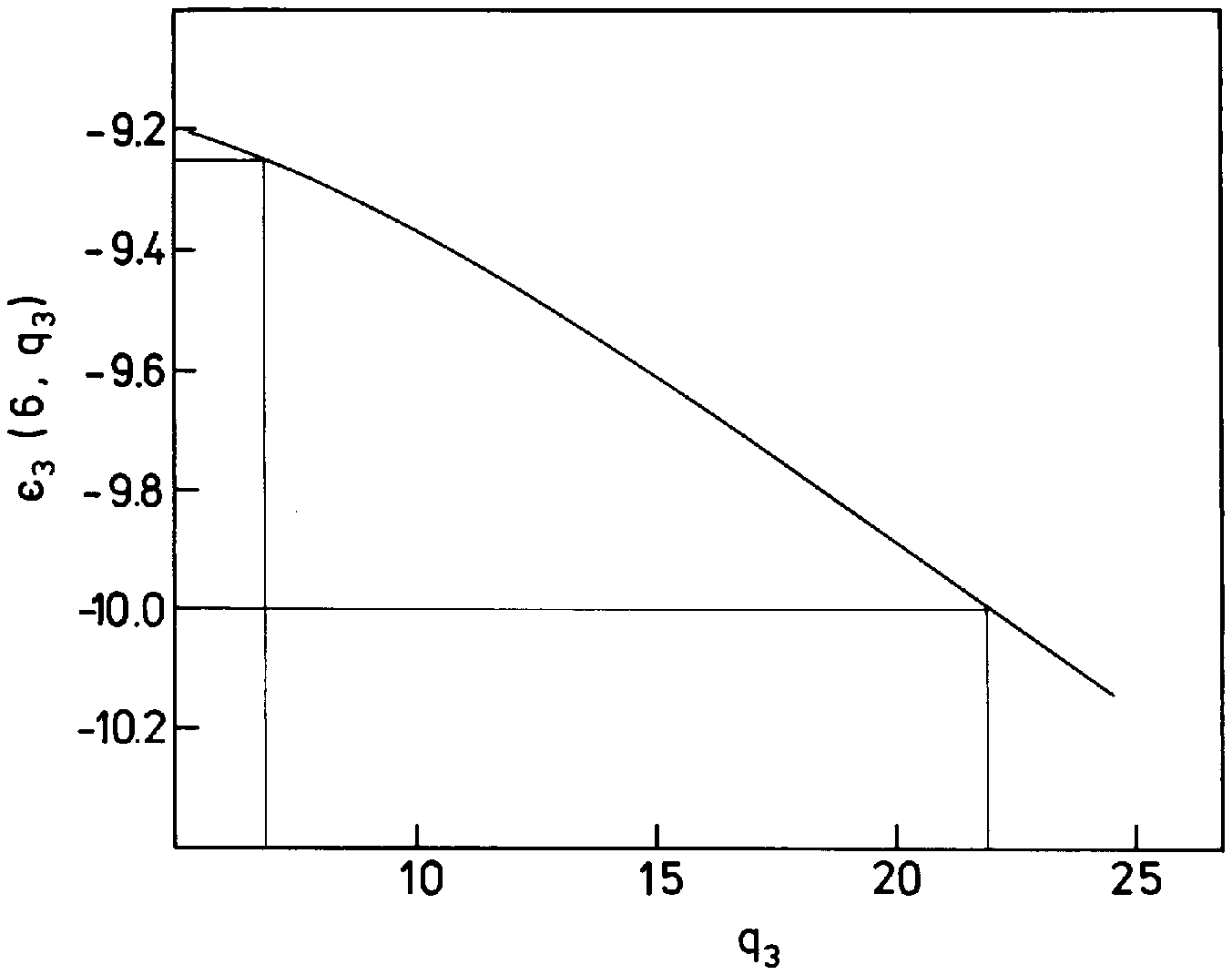}
\epsfxsize=7cm \epsfbox{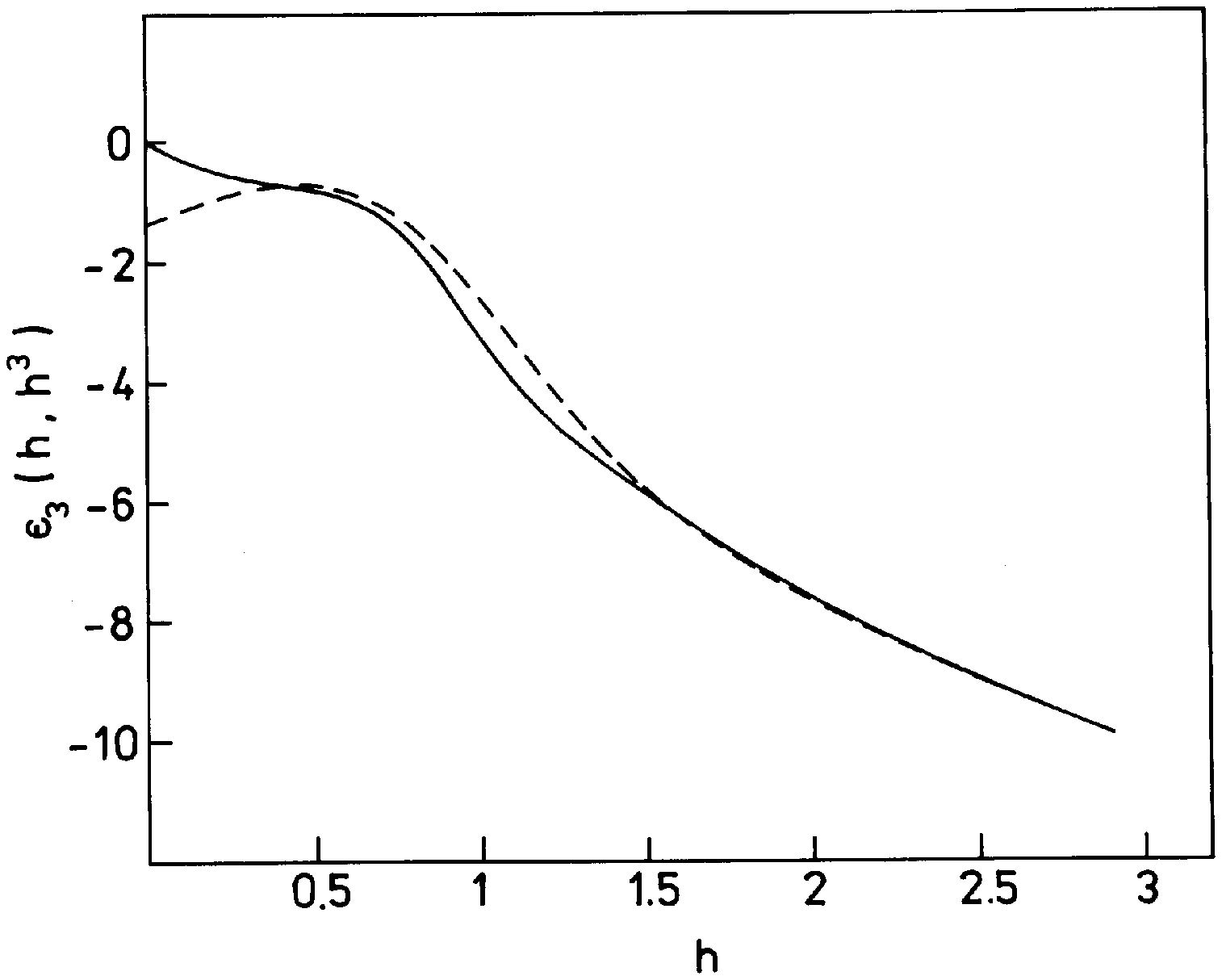}
\vspace*{-16.25cm}\hspace*{6cm} {\em a} \vspace*{4.75cm} \newline 
 \hspace*{6cm} {\em b} \vspace*{4.75cm} \newline
 \hspace*{6cm} {\em c} 
\vspace*{5cm}
\caption{ Holomorphic energy $\protect\epsilon_3$ as the function of two
 continuous
parameters. (a) at fixed $q_3=2\protect\sqrt{3}$, (b) at fixed $h=6$ 
and (c) comparison
with the asymptotic formula (dashed line), Eq.(\ref{asym}), described in text.
}
\label{fig:f3}
\end{figure}

\begin{figure}[htb]
\vspace{9pt}
\epsfxsize=7cm \epsfbox{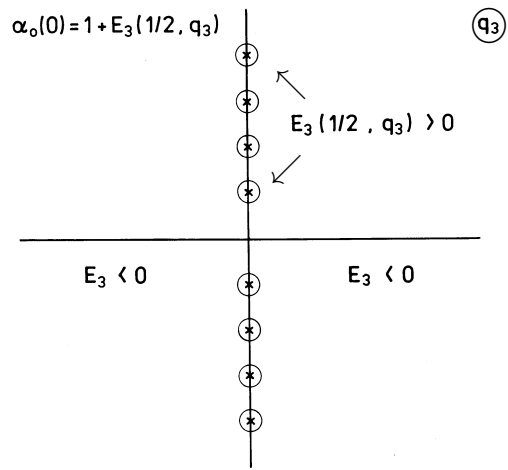}  
\vspace*{-5.8cm}\hspace*{3.8cm}$\nwarrow$ \vspace*{0.6cm}\newline
\hspace*{3.6cm} $\swarrow$ \vspace*{-2.4cm} \newline \hspace*{6.25cm} 
{\large $\bigcirc$}
\vspace*{8cm}
\caption{ Schematic map of the analyticity structure of $\protect\epsilon_3(1/2,q_3)$
in the complex $q_3$ plane. $E_3$ is positive only in the vicinity of the
poles. 
}
\label{fig:f4}
\end{figure}

\end{document}